# New abundances for old stars – atomic diffusion at work in NGC 6397

A homogeneous spectroscopic analysis of unevolved and evolved stars in the metal-poor globular cluster NGC 6397 with FLAMES-UVES reveals systematic trends of stellar surface abundances that are likely caused by atomic diffusion. This finding helps to understand, among other issues, why the lithium abundances of old halo stars are significantly lower than the abundance found to be produced shortly after the Big Bang.

Andreas Korn, Uppsala Astronomical Observatory, Sweden
Frank Grundahl. Århus University, Denmark
Olivier Richard, University of Montpellier II, France
Paul Barklem, Uppsala Astronomical Observatory, Sweden
Lyudmila, Mashonkina, Institute of Astronomy, Moscow, Russia
Remo Collet, Uppsala Astronomical Observatory, Sweden
Nikolai Piskunov, Uppsala Astronomical Observatory, Sweden
Bengt Gustafsson, Uppsala Astronomical Observatory, Sweden

When Joseph Chamberlain and Lawrence Aller in 1951 announced the discovery of stars significantly more metal-poor than the Sun, a new field of astronomical research was born: the observational study of nucleogenesis (or cosmochemistry) which attempts to answer the question of where the chemical elements come from and how their build-up proceeds with time. Today, while many details remain unsettled, we have a general understanding of how the cosmos developed chemically, from a mixture of hydrogen, helium and traces of lithium a few minutes after the Big Bang to stars as metal-rich as the Sun and beyond. The majority of this knowledge has been gathered by studying starlight by means of quantitative spectroscopy. Solar-type stars (here defined to be stars of spectral types F, G and K) have always played a central role in cosmochemical studies, primarily for two reasons: they have rich photospheric spectra (allowing a great variety of elements to be studied) and are long-lived (allowing all phases of Galactic chemical evolution to be investigated).

The art of deriving chemical abundances from spectra is to relate the observed line strengths of a certain element to its abundance in the stellar atmosphere these lines originate in. This is achieved by capturing the essence of the matter-light interaction in a theoretical model. To make this problem computationally feasible, a number of assumptions about the physics of stellar atmospheres are introduced: the atmosphere is assumed to be well-represented by a one-dimensional and static temperature and pressure structure in local thermodynamic equilibrium (LTE), convection is treated according to the mixing-length recipe, rotation, mass loss and magnetic fields are disregarded all together. All these are traditional assumptions, to mention only the most obvious ones. Some of these have in the meantime been abandoned, but many analyses still rest on them and are adorned with the flattering title “classical analysis”. One less explicit assumption often made when interpreting the abundance results concerns the chemical abundances themselves: they are assumed to be representative of the material the star originally formed out of. This means in particular that there are no physical processes which alter photospheric abundances with time.

It comes with the profession of a theorist to question the (sometimes bold) approximations made by more observationally inclined astrophysicists. Lawrence Aller, Evry Schatzmann and others addressed the problem of atomic diffusion already in the 1960s. Later, when a common abundance of lithium was found among warm halo stars by Monique and François Spite (1982; the so-called Spite plateau) and interpreted as a relic of the Big Bang, Georges Michaud and colleagues presented models of stellar evolution with atomic diffusion that “change the lithium abundance by at least a factor of about 2 in solar-type stars”. Larger effects were predicted for Population II stars. In other words, this study showed that lithium and other elements slowly settle into the star under the force of gravity. In particular for old stars the assumption of the constancy of photospheric abundances seemed questionable.

Models with atomic diffusion were subsequently shown to be very successful in reproducing the abundances of hot, chemically peculiar (CP) stars. The significance of diffusion for solar-type stars has, however, been seriously questioned: predicted effects from early models were quite large and failed to, e.g., meet the observational constraint of a flat and thin Spite plateau of lithium (Ryan et al. 1999). This problem has been alleviated in recent years by including additional effects like radiative levitation and turbulent mixing which counterbalance gravitational settling.

**Putting diffusion in solar-type stars to the test**

Globular clusters of the Galactic halo are primary testbeds for stellar evolution theory in general and for effects of atomic diffusion in particular. Stars in globular clusters have the same age and initial composition (with certain exceptions). To test the atomic-diffusion hypothesis, one thus compares photospheric abundances of stars at the main-sequence turnoff (where the effects of diffusion are largest) to the abundances of red giants (where the original heavy-element abundances are essentially restored due to the large radial extent of the outer convection zone). What sounds like a straightforward measurement in theory is challenging in practice: at a magnitude of $16.5^m$, turnoff stars in one of the most nearby globular clusters (NGC 6397, see Fig. 1 & 2) are too faint to analyse on 4m-class telescopes at high resolution and high signal-to-noise (S/N) ratio. It took the VLT and the efficient spectrograph UVES to analyse these stars for the first time.

When UVES became available in the late 1990s, there were two teams that focussed their efforts on NGC 6397. Raffaele Gratton and co-workers wanted to further constrain the nature of the anticorrelations of certain elements commonly found among giants by looking for them in unevolved stars. Frédéric Thévenin and colleagues investigated the connection between globular clusters and halo field stars as regards alpha-capture elements. Two different metallicities were advocated for the cluster by these two groups, log ε(Fe) = log ($N_{Fe}$/$N_H$) + 12 = 5.50 ± 0.01 (5 stars; Gratton et al. 2001) and log ε(Fe) = 5.23 ± 0.01 (7 stars, assuming LTE, Thévenin et al. 2001), where the error given is the standard deviation of the mean. Obviously, these results are quite incompatible with one another. Gratton et al. backed up their high metallicity scale by analysing three stars at the base of the red-giant branch (RGB) which gave log ε(Fe)$_{LTE}$ = 5.47 ± 0.03 suggesting the good agreement between the two groups of stars to be "a constraint on the impact of diffusion". But what if Thévenin et al. were right with their turnoff-star metallicity?

We started to investigate this issue in mid-2003 and uncovered potential systematic effects in the temperature determination of the Gratton et al. analysis which could produce the apparent abundance differences between the two analyses and mimic the good agreement between the turnoff and base-RGB stars. The problems found concern the imperfect removal of blaze residuals in the UVES pipeline spectra that Gratton et al. used to set constraints on the effective temperatures of the stars from the Balmer line Hα (cf. Korn 2002). With the help of sophisticated echelle-data reduction routines developed by Nikolai Piskunov and Jeff Valenti, we could show that the systematic corrections are non-negligible amounting to −250 K. We subsequently applied for observing time with FLAMES-UVES to investigate this issue further. Fibre-fed spectrographs are less prone to the above-mentioned problems and this multi-object facility seemed ideally suited for this research.

The proposal was accepted, but was moved to Visitor Mode, as we had requested the "old" (now "B") high-resolution settings for GIRAFFE. The visit to Paranal in June of 2004 was fruitless: high winds and thick clouds did not allow us to collect more than 20% of the necessary data. We continued the analysis of the archival data and in a talk at the ESO-Arcetri workshop on "Chemical Abundances and Mixing in Stars" in September 2004 concluded that "gravitational settling of iron of up to 0.1 dex at [Fe/H] ≈ −2 seems possible".

**A second chance with FLAMES**

The successful reapplication for observing time in Period 75 gave us a second chance, this time in Service Mode. We observed a variety of stars in NGC 6397, from the turnoff to the red-giant branch. In every observing block, two UVES fibres were given to two stars in the middle of the subgiant branch (SGB), one fibre monitored the sky background. Five turnoff stars were observed for a total of 12 hours, while the six RGB stars only required 1.5 h to reach a S/N of 100. With a total integration time of 18 h, the two SGB stars have the highest S/N. The 130 fibres to GIRAFFE were filled with stars along the SGB. After receiving the data, the analysis could begin, first with a look to the effective temperatures of the turnoff stars. Already at this early stage, we learned that our preliminary analysis essentially pointed in the right direction.

The fully spectroscopic analysis of the 18 FLAMES-UVES targets took a few months. Meaningful results were only obtained, once we had properly accounted for the sky background (via the sky fibre) and the fibre-to-fibre throughput correction. By November 2005 the spectroscopic results were ready and we asked Olivier Richard to compute diffusion models including radiative accelerations and turbulent mixing for comparison with our derived abundances. It is mainly the overall size of metal diffusion (in terms of iron) and the behaviour of calcium which set limits on the unknown strength of turbulent mixing (see Fig. 3). The analysis is based on traditional 1D models (see above), but for iron and calcium detailed (non-LTE) line formation was used. We also investigated the impact of using hydrodynamic model atmospheres and found very similar results for weak lines.

The main challenge lies in determining a realistic effective-temperature *difference* between the turnoff and the RGB stars. This is because most spectral lines originate from neutral species which react most strongly to the temperature. Besides, the surface-gravity difference is well-constrained by the apparent-magnitude difference of stars at a given distance.

For the spectroscopic analysis, we rely on Hα which gives an effective-temperature difference between turnoff and RGB stars of 1124 K (see Table 1). This number is in excellent agreement with the photometric estimate based on the Strömgren index ($v$−$y$) which indicates 1108 K. The surface-gravity differences agree equally well, the (logarithmic) spectroscopic value being 0.05 dex smaller than the photometric one. This good agreement between two independent techniques gave us confidence that we could determine relative abundance *differences* with high accuracy.

**Table 1** Mean stellar parameters of the FLAMES-UVES stars. Typical errors on $T_{eff}$ and log $g$ are 150 K and 0.15, respectively. The errors in log ε (Fe) are the combined values of the line-to-line scatter of Fe I and Fe II for the individual stars propagated into the mean value for the group.

| stars | # | $T_{eff}$ [K] | log ($g$ [cm/s$^2$]) | log ε (Fe) |
|---|---|---|---|---|
| turnoff | 5 | 6254 | 3.89 | 5.23 ± 0.04 |
| subgiant | 2 | 5805 | 3.58 | 5.27 ± 0.05 |
| base-RGB | 5 | 5456 | 3.37 | 5.33 ± 0.03 |
| RGB | 6 | 5130 | 2.56 | 5.39 ± 0.02 |

**The diffusion signature**

We uncovered systematic trends of abundances with evolutionary phase. The best determined abundance is that of iron for which the analysis rests on 20-40 lines of Fe I and Fe II. The abundance difference between turnoff and RGB stars is found to be (0.16 ± 0.05) dex (see Fig. 7) which is significant at the 3σ level. Other elements were also investigated (calcium and titanium). The trends for these elements are found to be shallower than for iron which is a specific prediction of the diffusion model with turbulent mixing (see Fig. 4 and 5).

This is the first time that metal diffusion in old stars is constrained by means of observations. Via the diffusion model, this also constrains helium diffusion. Comparison with a 13.5 Gyr isochrone constructed from the diffusion model with turbulent mixing shows that the spectroscopic stellar parameters meet the cosmological age constraint (see Fig. 6). Residual discrepancies are small, but may point towards a somewhat younger age for NGC 6397. As can be seen in Fig. 3, the absolute age assumed has only a minor impact on the results.

While helium diffusion is nowadays considered a standard ingredient in stellar-evolution theory, this is not the case for metal diffusion. The latter has, however, a smaller impact on stellar evolution of metal-poor stars and we therefore expect only moderate changes to isochrone ages of globular clusters in general (nonetheless, globular clusters are now more intriguing objects than ever, see Fig. 7). When it comes to unevolved field stars, the situation is different: the high ages for halo field stars that are reported in the literature are systematically overestimated by underestimated metallicities. Together with helium diffusion, the removal of this bias will likely bring down field-star isochrone ages by several billion years.

**The diffusion of lithium and its cosmological implications**

The recent upward revision of the baryonic matter content ($\Omega_b$) of the Universe by experiments like WMAP (see Spergel et al. 2006 for the most recent results) has led to revised abundances of deuterium, helium and lithium produced in Big-Bang nucleosynthesis (BBN). More lithium than previously thought is found to be produced in BBN ($\log \varepsilon(\mathrm{Li})_{\mathrm{BBN}} = 2.64 \pm 0.03$, “CMB+BBN” in Fig. 5). This has led to a distressing difference of about a factor of two or more with respect to the common abundance measured in stars on the Spite plateau. We now find that, once atomic diffusion is accounted for, the stellar abundances ($\log \varepsilon(\mathrm{Li}) = 2.54 \pm 0.10$) are in good agreement with the primordial lithium abundance (Fig. 5). Note that we even seem to see the effects of atomic diffusion in the behaviour of lithium in the turnoff and SGB stars: the measured upturn towards the SGB stars (before dilution sets in) is a specific prediction of the diffusion model.

With this work, the existence of the Spite plateau is more astonishing than ever: it is essentially of cosmological origin, but is *uniformly* lowered by physical processes in the stars. This theoretical idea has been around for decades, yet it never really caught on with the observers. Only recently, with the latest diffusion models making such a uniform lowering plausible and the WMAP results constraining the primordial lithium abundance, has the idea gained credence. Atomic diffusion may thus resolve the cosmological lithium discrepancy (Korn et al. 2006), but we caution that other effects (the absolute temperature scale of Population II stars and the related issue of a trend of lithium with metallicity, the chemical evolution of lithium in the early Galaxy, see Asplund et al. 2006 for a recent review) may also play a significant role.

This work highlights three methodological points: the possibilities of the new generation of multi-object spectrometers at the largest telescopes, the virtue of a homogeneous and consistent analysis of all relevant criteria (spectroscopic as well as photometric) and the benefits of developing physical models to a higher degree of self-consistency. The empirical result that atomic diffusion affects atmospheric chemical abundances of old unevolved stars and that models can account for its effects is significant and has a number of consequences, some of which were discussed above. For the time being, turbulent mixing is introduced in an ad-hoc manner, without specifying the physics behind it. Likewise, we neglect the spectral effects of the thermal inhomogeneities and the 3D nature of convection in our traditional 1D analysis. Thus, a lot remains to be done before the quality of current observational data is adequately matched by physical models of stars, so that we can fully read the fingerprints of chemical elements in stellar spectra.

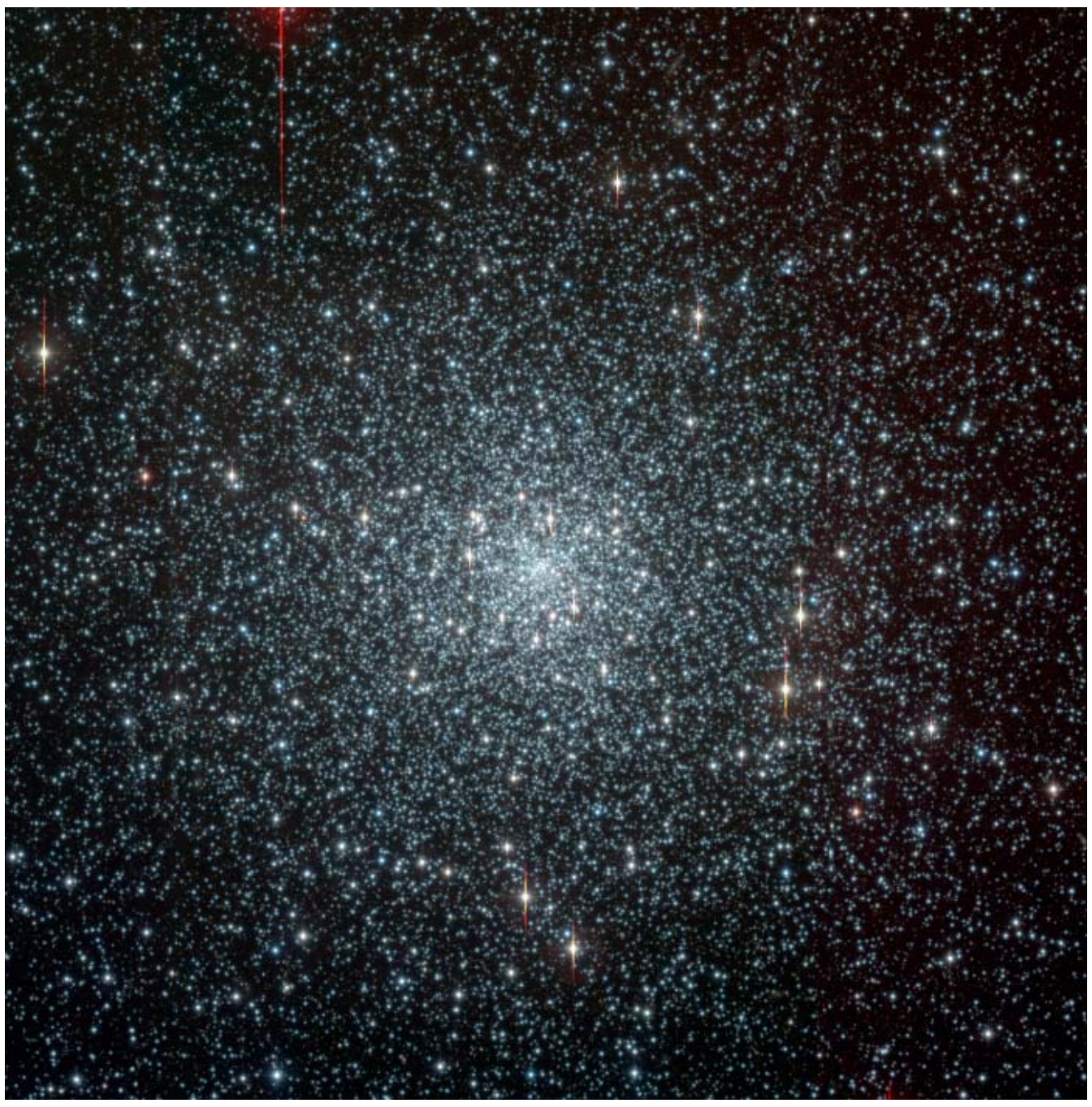

**Figure 1** The nearby metal-poor globular cluster NGC 6397 as seen by the Wide-Field Imager on the ESO/MPI 2.2m. Its distance modulus is $(m - M) \approx 12.4^m$.

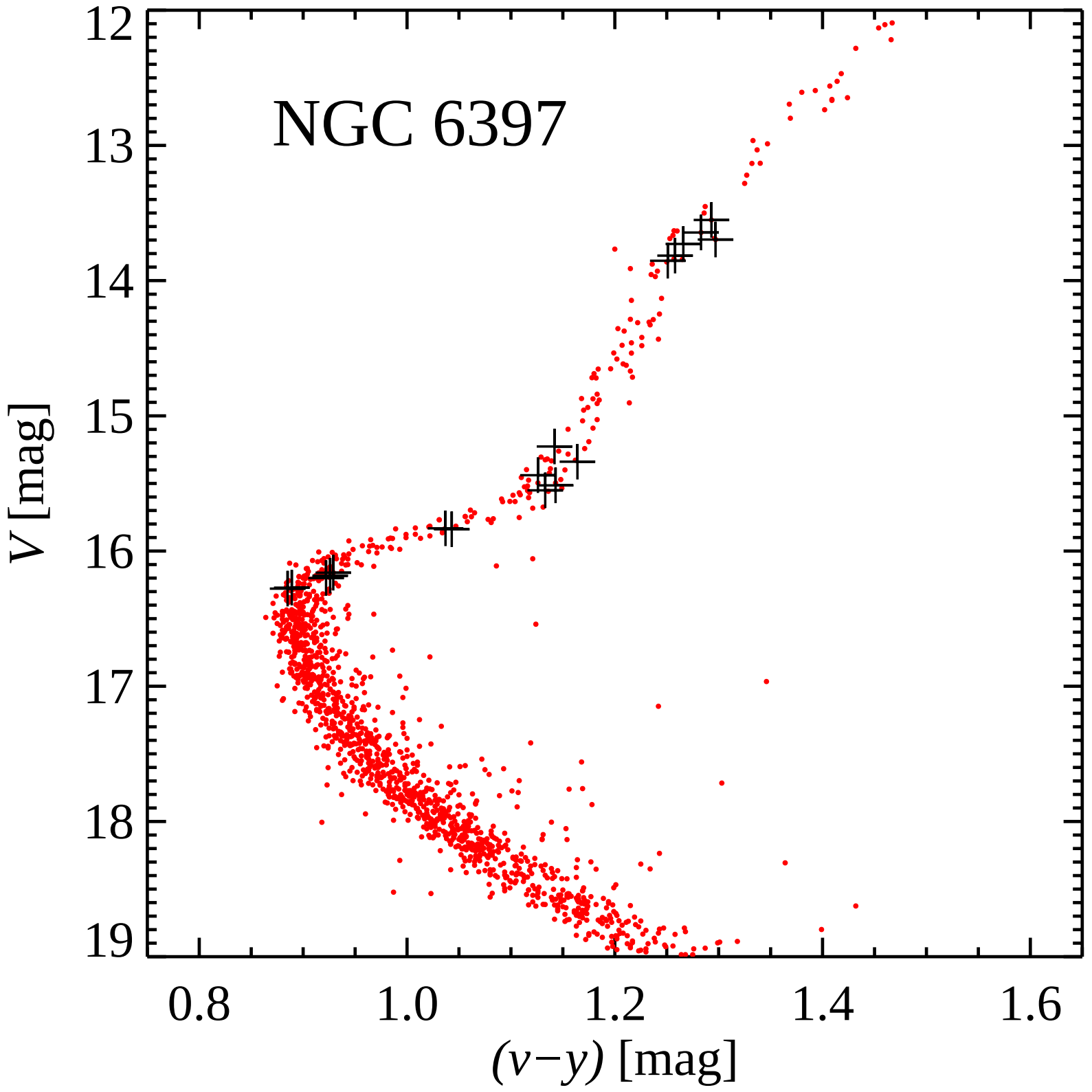

**Figure 2** Colour-magnitude diagram of NGC 6397 with the FLAMES-UVES targets marked by the crosses. According to literature values, the metallicity of this cluster is just below 1/100th solar ([Fe/H] ≈ −2.1). The data was acquired with the Danish 1.54m telescope on LaSilla.

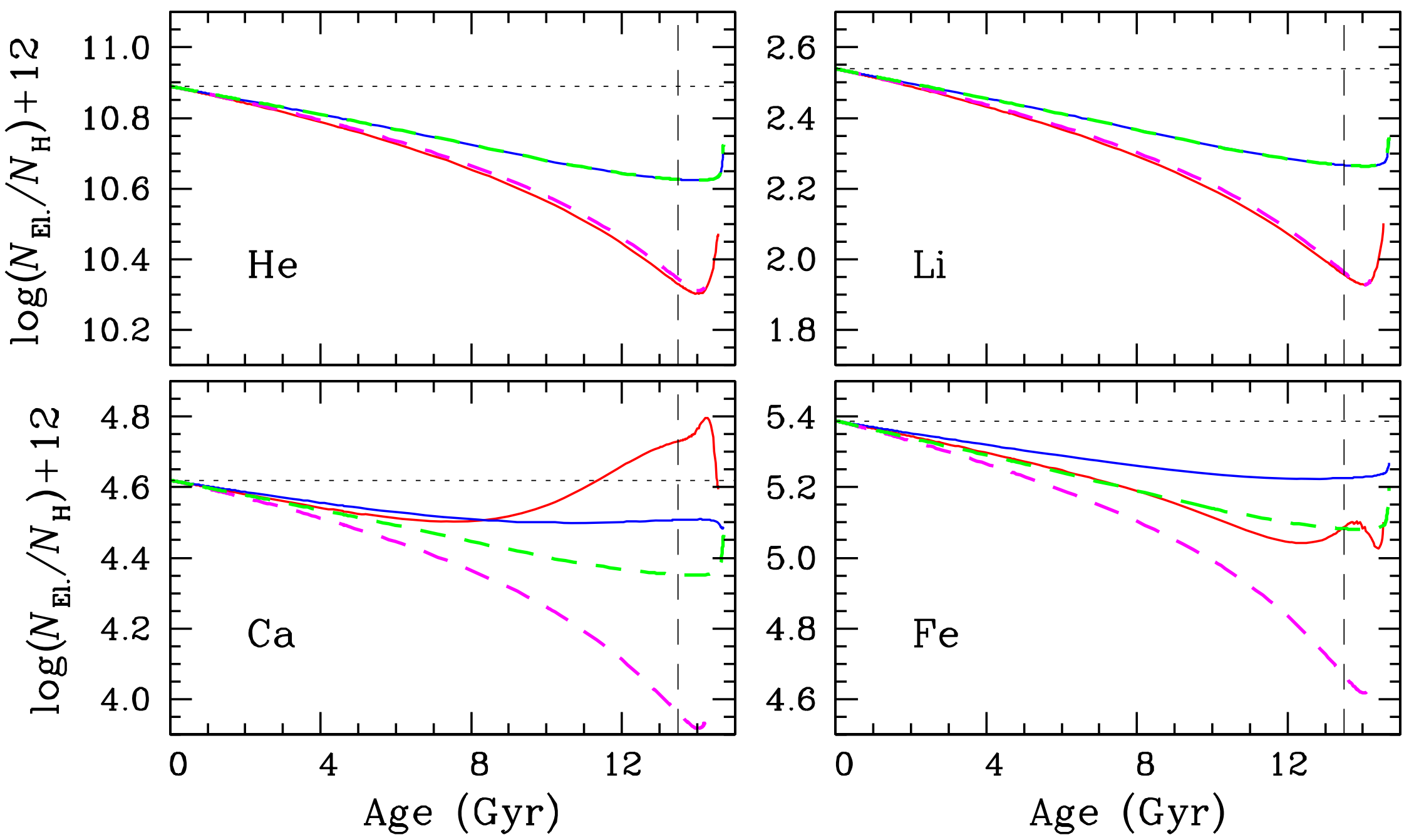

**Figure 3** Theoretical predictions for a 0.77 $M_{solar}$, [Fe/H] = −2 star (located at the turnoff after 13.5 Gyr) as a function of age using different input physics: with atomic diffusion only (dashed magenta), with atomic diffusion and radiative levitation (red), with atomic diffusion and turbulent mixing, but without radiative levitation (dashed green), with atomic diffusion, turbulent mixing and radiative levitation (blue). The dashed horizontal line indicates the original abundance. The panels for Ca and Fe clearly show the importance of radiative levitation for heavy elements. The blue model describes our observations of calcium and iron best.

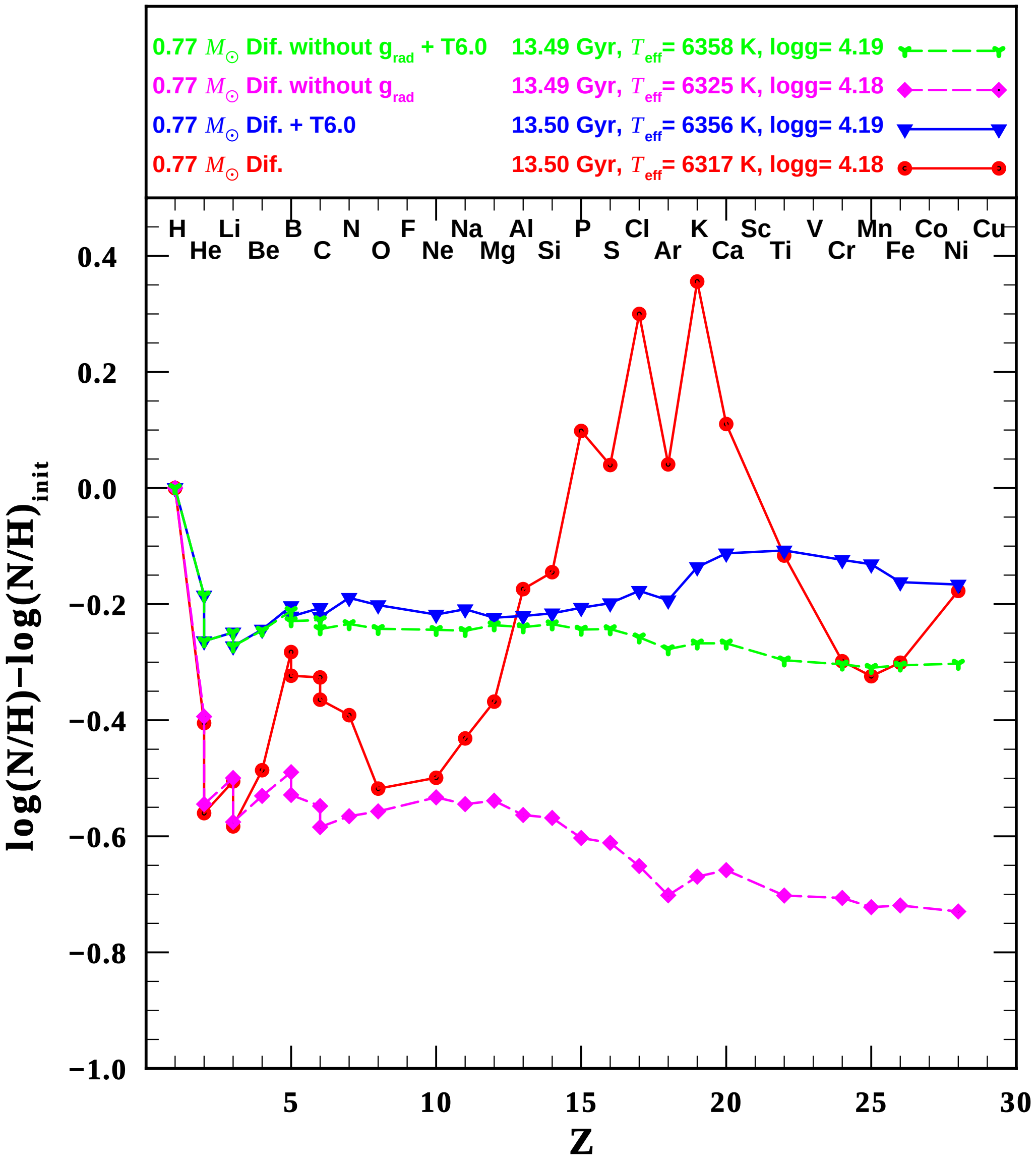

**Figure 4** Abundance variations for a 0.77 $M_{solar}$, [Fe/H] = −2 star after 13.5 Gyr. As in Fig. 3, the model describing our observations is the blue one. Among the heavy elements (Z ≥ 20), the abundance change is minimal for calcium and titanium, while it is largest for iron and nickel. Lighter elements show even larger variations (e.g. magnesium and aluminium), but are subject to anticorrelations in globular clusters (Gratton et al. 2001). Of all elements, helium and lithium are affected the most.

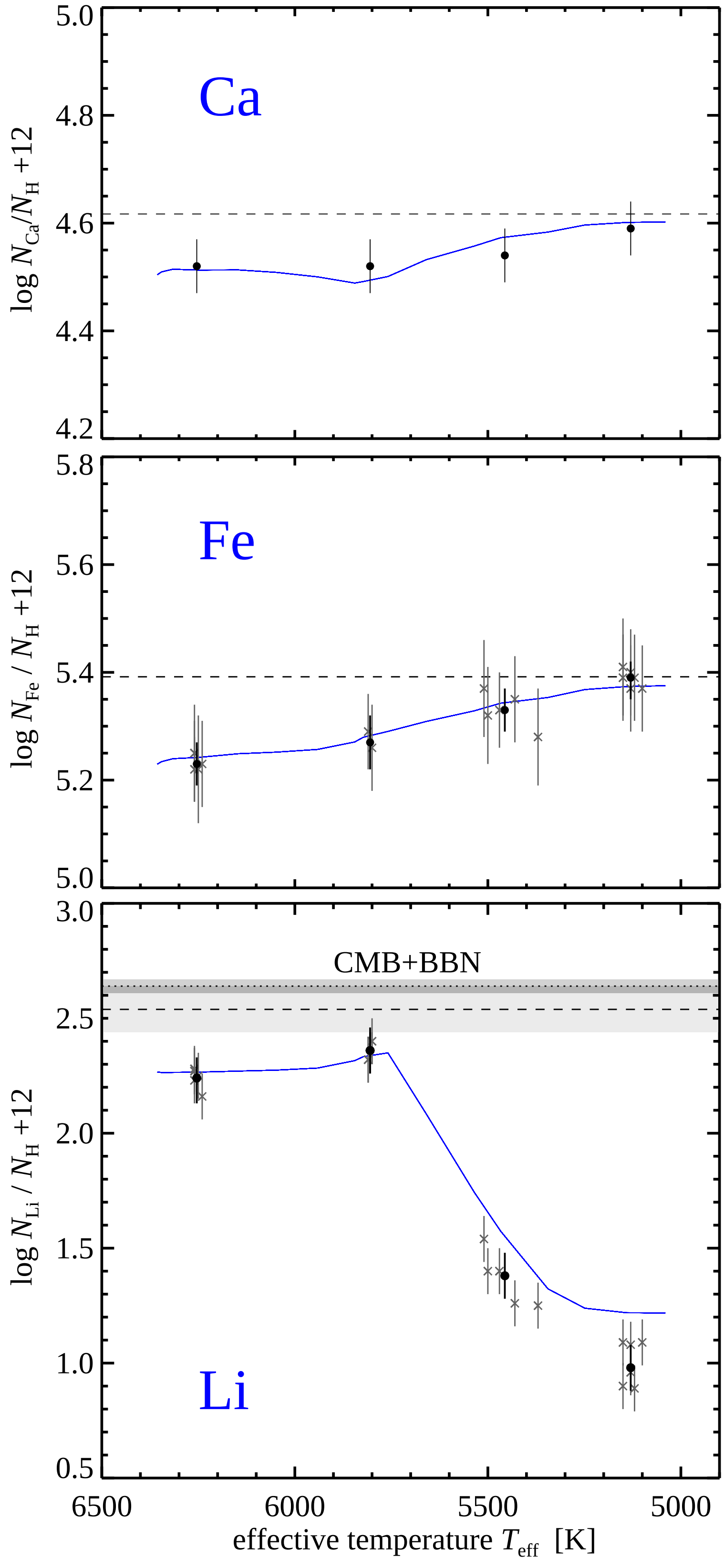

**Figure 5** Observed trends of abundance with effective temperature for calcium, iron and lithium. The blue line represents the prediction from the diffusion model with turbulent mixing at an age of 13.5 Gyr (see Fig. 3) with the dashed line indicating the original abundance. Measurements for individual stars are marked by crosses, group averages by bullets. For calcium, measurements were made on the group-averaged spectra to increase the S/N.

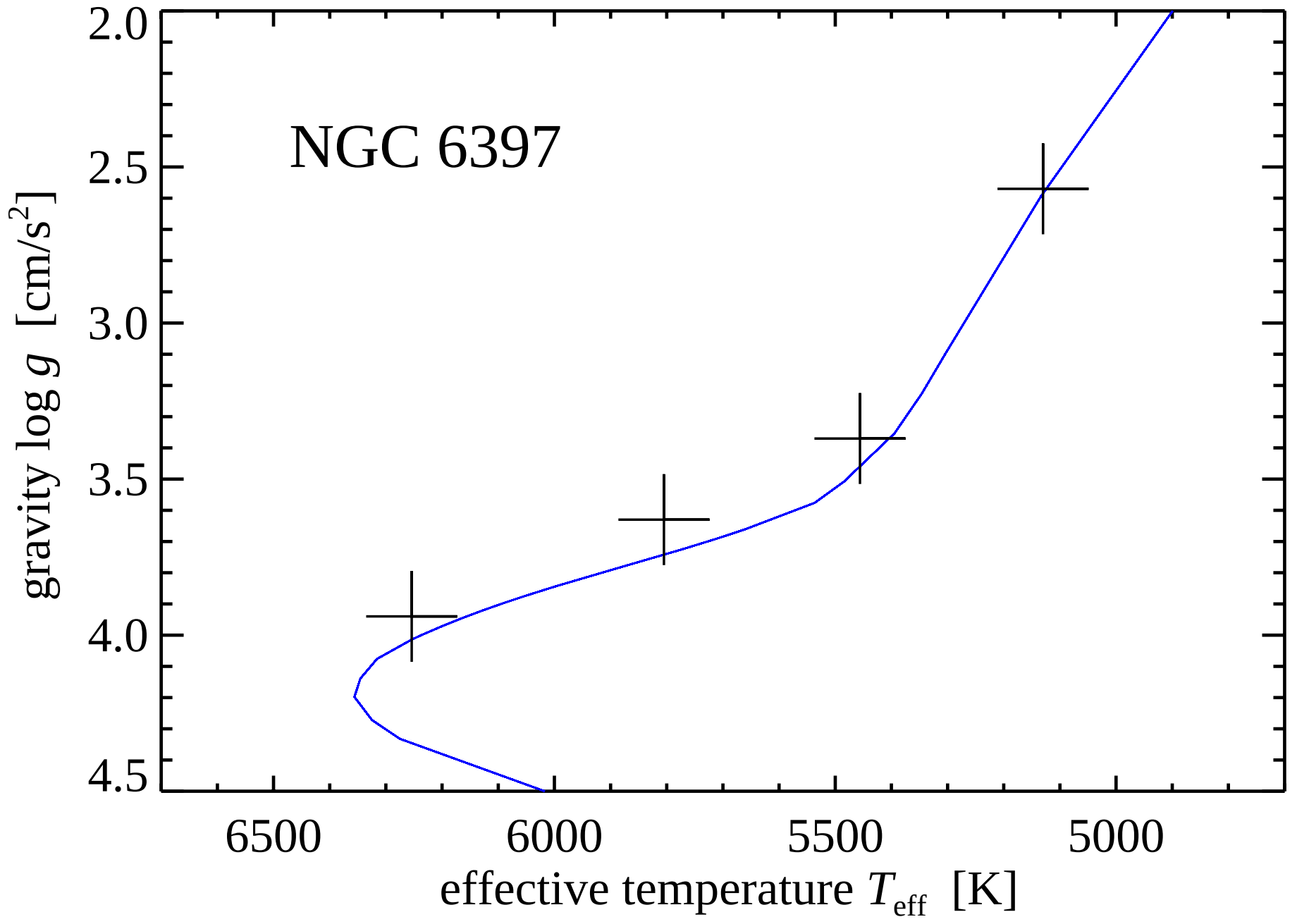

**Figure 6** Kiel diagram showing the mean positions of the stars relative to a 13.5 Gyr isochrone constructed from the diffusion model with turbulent mixing. Unlike in Table 1, helium diffusion has been considered as a structural effect in the spectroscopic analysis which results in higher gravities (by +0.05 dex) for the turnoff and SGB stars (see Korn et al. 2006 for details).

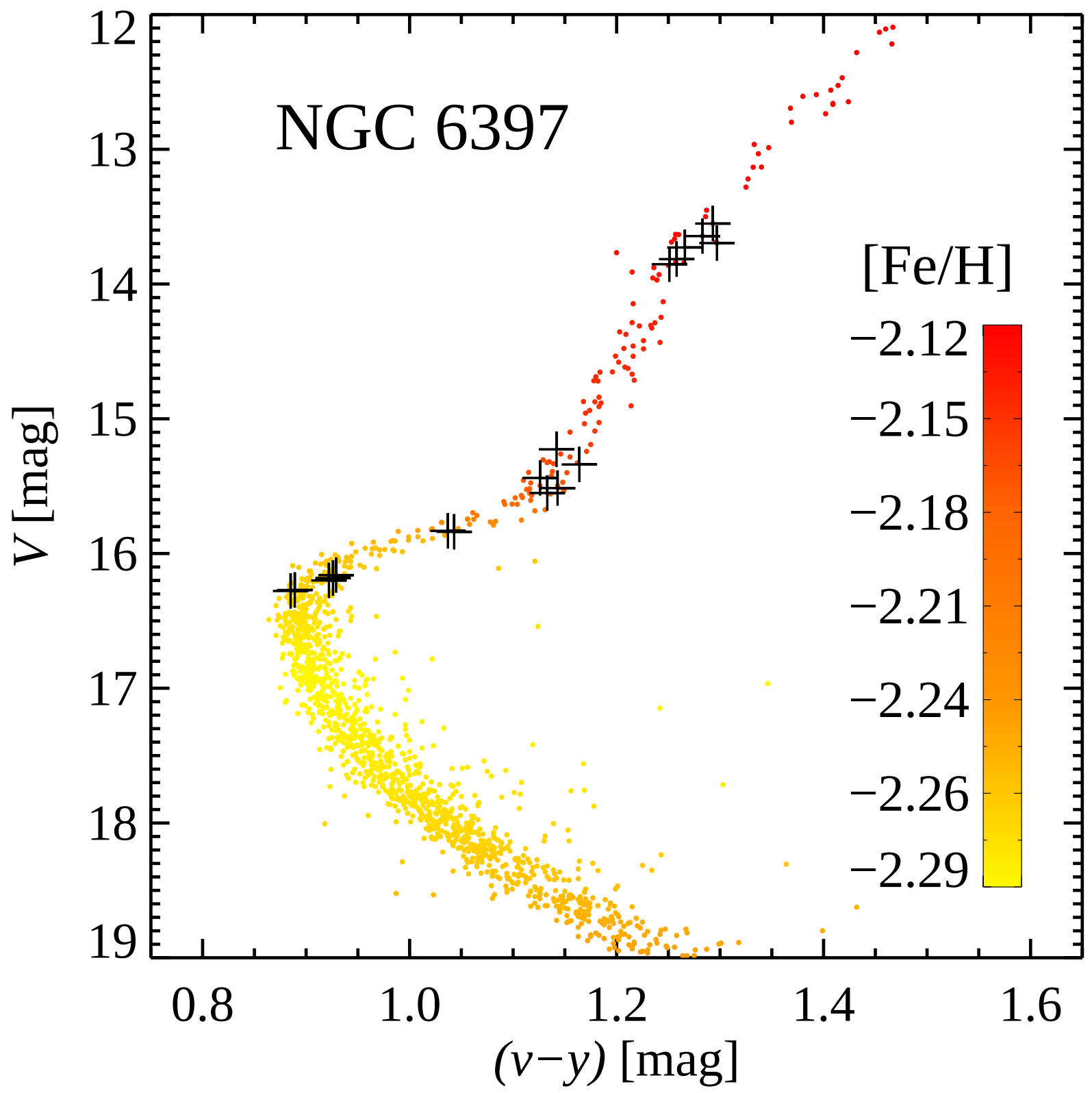

**Figure 7** Colour-magnitude diagram of NGC 6397 with the FLAMES-UVES targets marked by the crosses. The colour-coding represents abundance trends in iron taken from the diffusion model with turbulent mixing calibrated on the measured abundance trends. All stars were once born with an initial iron abundance around [Fe/H] = −2.12, but unevolved stars now display lower abundances in their atmospheres.